# Generation and Balancing Capacity in Future Electric Power Systems
# - Scenario Analysis Using Bayesian Networks

Seppo Borenius [a*], Pekka Kekolahti [a], Petri Mähönen[a], Matti Lehtonen[a]

[a] Aalto University, School of Electrical Engineering, PO Box15500, FI-00076 AALTO, Finland
[*] Corresponding author, email addresses: seppo.borenius@aalto.fi

**Abstract**

This paper examines the evolution of the Finnish electric energy system up to 2035, focusing on the likelihood of different development paths. The primary contribution of this paper is the development of an extensive Bayesian Network, designed to model and analyse the evolution of power generation capacity mix, assess the likelihood of different grid management scenarios, and understand the causal relationships underlying these scenarios. A target optimisation was carried out using the constructed Bayesian Network to explore possibilities to minimise grid management complexity. The results of the optimisation reveal that the authorities and stakeholders should prioritise increasing demand response, gas power, and battery storage capacities. These mature technologies are well-suited to guarantee energy adequacy during peak consumption periods, which in Finland typically occur during consecutive cold, dark and windless winter weeks. Although this study focuses on the evolution of the Finnish power grid, the constructed Bayesian Network approach is broadly applicable and can be utilised to explore causal relationships in other countries by employing the designed questionnaire and engaging a panel of experts specific to the country's energy infrastructure.

**Keywords** Power generation mix; Grid management; Bayesian Networks; Causality analysis; Scenario planning

## 1. INTRODUCTION

The Finnish electric energy system is undergoing significant changes in order to address the challenges of climate change and to achieve carbon neutrality. Key enablers in this transition include increased electrification and the evolution of the power grid to integrate large amounts of renewable energy generation, paving the way towards a sustainable, carbon-neutral energy system. Therefore, to ensure optimal investments, reliability, and cost efficiency, it is crucial to explore alternative evolution pathways for the electric power system. Several prominent organisations have published electrification scenarios for Finland with timelines extending to 2035, 2045, and 2050, aiming to assess potential strategies for the power generation [1],[2],[3]. These scenarios encompass diverse trajectories for energy production and consumption, with substantial variation in estimated production and usage patterns.

Previous research by some of this study's authors has analysed the evolution of the Finnish power grid up to 2035 from a grid management perspective [4]. This research employed a formal scenario planning process and insights from a senior expert panel to create four *grid management scenarios* [4],[5]. These *scenarios* were structured by examining two key uncertainties: (1) "*Domestic controllable bulk generation capacity* increasing?" and (2) "*New cost-effective large-scale domestic balancing power solutions emerging?*". As shown in Fig. 1, this framework yielded four distinct *grid management scenarios*: Scenario B1 *Top-down grid*, Scenario B2 *Decentrally balanced grid*, Scenario B3 *Centrally balanced grid*, and Scenario B4 *Highly distributed grid*.

In Scenario B1, *the Top-down grid*, domestic controllable bulk generation and domestic balancing power capacity are roughly adequate reducing the need for large-scale demand response capacity. Scenario B1 resembles the traditional situation where power generation primarily from large controllable synchronous generators adjust in real-time to match demand and the electricity flow towards the customers. Scenario B4, *the Highly distributed grid,* represents the opposite case, requiring significant imports of both bulk and balancing power to maintain power balance and grid stability. Scenario B2, *the Decentrally balanced grid*, represents the case where domestic controllable bulk generation would be roughly sufficient; however, large-scale balancing power would be inadequate. This scenario would require extensive distributed balancing solutions such as demand response, to substitute for condensing power plants and other controllable generation that could be phased out. *Scenario B3, the Centrally balanced grid*, assumes roughly sufficient domestic balancing power, but insufficient domestic bulk generation capacity. In this case, frequent balancing actions would be needed due to the lack of controllable bulk generation capacity. Nevertheless, Scenario B3 would be comparatively easily manageable relative to Scenario B2, as it would not require aggregating balancing power from numerous small resources. For further details on these *grid management scenarios*, we refer readers to Ref. [4]. It should be noted that while developing the *grid management scenarios*, the expert panel assumed that the import capacity would remain limited, and that reasonable self-sufficiency would be necessary to ensure adequate security of supply during crisis.

|  | Domestic controllable bulk generation capacity increasing ? | |
|---|---|---|
|  | Yes | No |
| **A new, cost-effective domestic large-scale balancing power solution emerging?** No | B2 Decentrally balanced grid | B4 Highly distributed grid |
| Yes | B1 Top-down grid | B3 Centrally balanced grid |

Fig. 1. The considered *grid management scenarios* [4]

All the above mentioned scenarios ([1],[2],[3][4]) were developed without assumptions regarding their likelihoods or considerations of how the power generation capacity mix might affect the controllability of the power balance and the ability to maintain an stable, adequate, and continuous energy supply. This paper contributes by constructing a four-layer hierarchical Bayesian Network (BN) to document expert opinions on power grid evolution and by expanding the analysis of the four *grid management scenarios* by estimating their respective likelihoods. Additionally, it systematic assesses the power generation capacity mix within each scenario, utilising quantitative metrics to enable a more detailed evaluation.

The research problem is stated as follows: How do various factors – ranging from technical and economic aspects to power generation, storage, and control methods – shape the probability of different *grid management scenarios* when the import of electricity is restricted? We aim to answer the following research questions:

- RQ1: What is the estimated power generation capacity mix in Finland in 2035, and how is it influenced by various technical, commercial, socio-economic, and political factors?
- RQ2: How does the interplay between diverse electricity generation, advanced storage capacities, and innovative demand response potential shape the probability of *grid management scenarios* in 2035?
- RQ3: How can Bayesian Networks be constructed based on expert input to capture the dependences related to RQ1 and RQ2, enabling both inductive and deductive analysis?

To the best of our knowledge, the used approach is unique in two ways. First, it estimates the probabilities of different *grid management scenarios* based on the power generation, power storage, and demand response capacities. Second, it employs a Bayesian Networks approach to model these dependencies. This innovation enables the utilisation of the constructed Bayesian network for deductive, inductive and sensitivity analysis.

The rest of the paper is organised as follows. Chapter 2 provides a brief overview of the methods, and Chapter 3 describes their application in this study. Chapter 4 presents the results divided into three major categories. Finally, Chapter 5 provides discussion and concluding remarks.

## 2. METHODOLOGY OF STUDY

In this study we investigate factors influencing *domestic controllable bulk generation capacity* and *new cost-effective domestic large-scale balancing power capacity*, as well as the likelihood of various *grid management scenarios* (cf. terminology presented in Chapter 1 and Fig. 1). Apart of collecting data from literature review, we have developed and conducted an expert survey to estimate power generation capacities for 2035. As a major quantitative method we utilise Bayesian Network (BN) modelling to analyse the collected data.

The expert survey gathers knowledge through interviews with 15 senior industry experts and a structured set of questions. These experts are seasoned academic and industry professionals with deep expertise in the power grid domain, including future scenarios for Finnish power grids. A common approach for knowledge elicitation is a group work with multiple variations, which can facilitate consensus among the survey panel [6] as well as approaches, such as Delphi-method, RAND/UCLA Appropriateness Method (RAM) and Nominal Group Technique (NGT) [7],[8]. Among the methods listed above, this study employs a specialised approach that most closely aligns with the RAM.

Different expert weighting schemes are utilised to increase the accuracy of estimations [9]. When constructing a Bayesian Networks (BN) from experts' opinions, as is done in this study, likelihood weighting is suggested by Stiber et al. [10] and Farr et al. [11]. Furthermore, Farr et al. [11] propose Prior Linear Pooling-method for weighting, when all experts share a common Bayesian Network, as in this study.

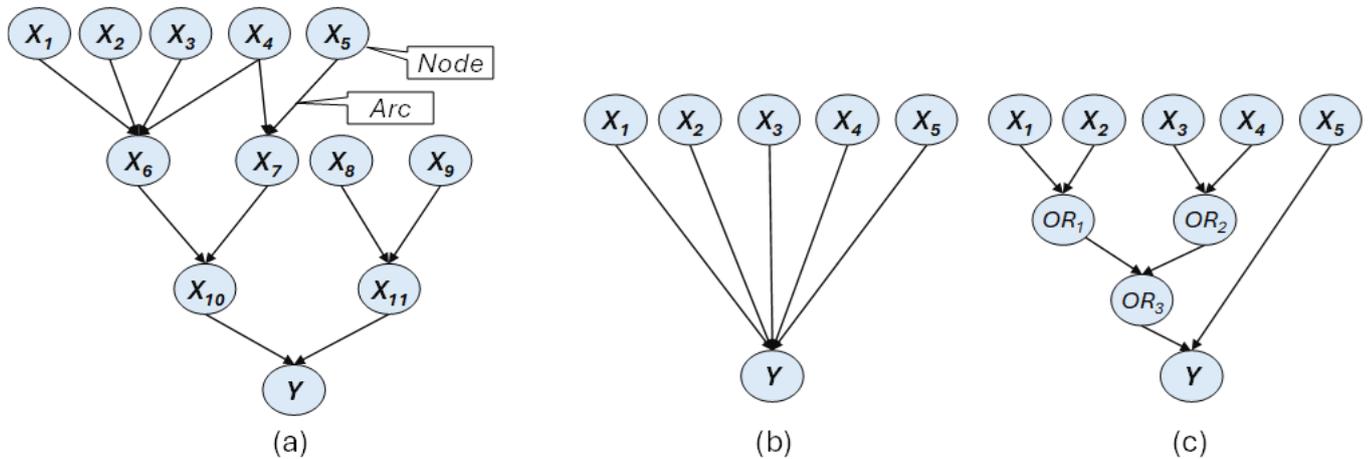

Fig. 2. (a) An example of a Bayesian Network with a 4-layer hierarchy, where $X_{10}$ and $X_{11}$ are direct and $X_1 – X_9$ indirect causes for the effect $Y$; (b) an example of a Bayesian Network with 5 direct causes for $Y$ and (c) the same but using 'parent friendly divorcing'.

Bayesian statistics enables the updating (revisioning) of existing prior probabilities (beliefs) as new information (evidence) is gained. The updated probabilities are called *'a posteriori'* probabilities and the prior probabilities *'a priori'* probabilities. A Bayesian Network is a probabilistic graphical model where random variables, represented as nodes ($X_i$) in the BN, are interconnected by arcs that delineate the conditional dependencies between these nodes as depicted in Fig. 2 [12], [13], [14]. An arc represents either a statistical or causal dependency between the nodes. For the purposes of this study, a pre-defined four-layer hierarchical structure (cf. Fig. 2a) has been chosen for the causal Bayesian Network model. Firstly, this approach is due to the hierarchical model's capability to represent complex relationships and its structured approach for handling various levels of abstraction and distinct stages in the modelling process. Secondly, the predefined structure is more practical due to the experts' limited time available for the interviews. Thirdly, a carefully defined BN structure minimises the number of the questions that need to be asked from the experts.

When the number of parent nodes of a child increases, the Conditional Probability Table (CPT) grows exponentially. This may lead to situations where the extensive size of a CPT results in an overwhelming number of assessment questions for experts, decreasing their motivation to respond. Several methods have been devised to assist in the populating CPTs in Bayesian Networks. These include the weighted sum algorithm introduced by Das [15], the ranked node algorithm of Fenton et al. [16] and algorithms like Noisy-OR and its derivates (called here Noisy-X) [17], [18], [19], [20], [21]. Noisy-X methods fit into modelling complex causal interactions, where they can form a reasonably good real-world approximation. They have a characteristic called Independence of Causal Influence (ICI), which means that the parent nodes have a certain amount of independence with each other in the way they affect the child. The number of assessment questions using Noisy-X increase linearly with the number of parents rather than exponentially, as in the general CPT-case.

The Noisy-OR -approach can be used when there are multiple potential causes, each of which can independently cause the outcome to occur, but only with a specific probability. Given the binary child node $Y$ and its binary parents $X_1, X_2, X_3, ..., X_n$, and the following definitions:
- $\theta_i$ : the probability that the $i$-th cause $X_i$, when present (i.e., $X_i =1$), independently causes the effect $Y$ to occur.
- $X_i$: degree of presence of $i$-th cause in Noisy-OR -context, where $0 \leq X_i \leq 1$. In this study $X_i$ is either 1 (cause is fully present) or 0 (cause is completely absent)
- $\lambda$ : the Leak probability containing unexplained variance related to $Y$
- $Y$: a binary variable that represents the occurrence of the effect. It takes a value of 1 if the effect occurs and 0 if it does not

then the generalised Noisy-OR is defined by the following equation (Eq. 1):

- Noisy-OR: $P(Y= 1| X_1, X_2, X_3, ..., X_n) = 1 - (1 - \lambda) * \prod_{i=1}^{n}(1 - \theta_i X_i)$  Eq. 1

This study assumes, that each parent node in the model can alone and independently cause the outcome node ($Y$) to be true, i.e., the study applies the Noisy-OR approach. This implies, for instance, that a specific component of *power generation capacity mix* can theoretically trigger the bulk capacity on its own, without the need for other components, allowing Eq. 1 to be applied. However, the computational effort required for large CPTs remains substantial. To address this issue, ICI implementations, alongside Noisy-X, often utilise a technique called 'parent friendly divorcing' [22] as described in Fig. 2b and 2c. This restructuring method introduces intermediary parents to maintain original conditional probabilities while simultaneously reducing computational complexity.

## 3. RESEARCH APPROACH

The literature review and analysis of the electrification scenarios devised in Finland for 2035 ([1],[2],[3]) serves two purposes. First, the data on power generation, storage, and demand response capacity, measured in gigawatts (GW), is provided as background information to the expert panel following the RAM principles. This serves as a foundation for the panel members to calibrate their opinions and assessments. Second, the overall *controllable domestic bulk generation power* capacity as well as the overall *new cost-effective domestic large-large balancing power* capacity in 2035 is calculated from the literature data and used as input for the triggering capacity values in the assessment questions for the experts. In this study, our primary aim is not to question the capacity values derived from the literature, but to understand the experts' beliefs and their confidence levels regarding the various components of power generation capacity mix and their peak capacities that may contribute to achieving the overall bulk and balancing capacities. This understanding, in turn, informs the probability of different *grid management scenarios* B1 – B4.

The expert assessment is divided into four areas based on the planned hierarchical structure of the Bayesian Network, depicted in Fig. 2a and Fig. 3:

- The first part, *Assessment Question Set 1a-c*, is to estimate the peak power generation capacity mix in gigawatts (GW) for 2035 at hierarchy layer L2, and to analyse how different external factors and circumstances at hierarchy layer L1 may affect these power generation capacity mix components.
- The second part, *Assessment Question Set 2*, is to estimate internal dependencies at layer L2, relevant for this study, i.e., how the available wind and solar power capacities in 2035 are used percentagewise for battery storage, P2X-X2P (Power-to-X, X-to-Power, referring to the process of converting electricity to hydrogen or its derivatives and then back to electricity), pumped hydro power (water power) storage, and direct electricity usage.
- The third part, *Assessment Question Set 3a-b*, is to estimate the experts' belief in how the selected components of the power generation capacity mix component at hierarchy layer L2 will affect the *new cost-effective domestic large-large balancing power* capacity, assuming its value will be greater or equal to 5 GW in 2035. Similarly, their task is to estimate how the selected power generation capacity mix components will affect the *controllable domestic bulk generation power* capacity, assuming its value will be greater or equal to 13 GW in 2035. Both capacities are situated at layer L3. The trigger values, i.e., the discretisation levels of 5 GW and 13 GW, were selected based on the findings from the literature review (Table 1) to approximately represent the mean from the 2035 scenarios and the 2023 values (from rows Mean Bulk and Balancing capacity and Current situation in Finland in Table 1). Using the values of Table 1, this would give (13,2+1,3)/2=7,25 and (13,8+12,9)/2=13,35 which were rounded 5GW and 13 GW, respectively. Regarding *new cost-effective domestic large-large balancing power* capacity somewhat lower value (5GW) than the arithmetic mean (7,25) was selected to ensure that the trigger value is *not* too high, which could lead to a situation were all experts' assessments fall below it.
- The fourth part, *Assessment Question Set 4*, is to estimate the effect of the *new cost-effective domestic large-large balancing power* capacity and *controllable domestic bulk generation power* capacity at layer L3 on the probability of the four *grid management scenarios*, which are situated at layer L4.

For more detailed description of the *Questions Sets* c.f. Appendix A.

The study employs a RAM-like approach for all four expert assessments, including utilising the literature review results to aid the panel in forming their opinions, encouraging experts to indicate their confidence levels in their estimations, and ensuring the number of experts (15) falls within the proposed range in RAM. To foster different views and to achieve broad distributions, the survey consisted only of one interview round. The goal was not to reach a consensus among the panellists, as the future remains inherently uncertain. Instead, the aim was to unveil different possible evolution paths and assess their likelihoods and causalities within them.

To help the experts to justify their probability values in *Question Sets 2* and *4* for assessing the CPTs and in estimation of the confidences, the authors provided them a nine-level dependency map between a typical description of the likelihood (impossible, almost no chance,…,certain) and its probability value (0%, 1-5%,…,100%) [23]. Similarly, a five-level Likert-scale was created to help in assessing causal strengths in *Question Set 1c* and *3*. The five levels of causal impact consisted of non-existent or very weak 0-20%, weak 21-40%, moderate 41-60%, strong 61-80%, very strong 81-100%.

Several commercial tools are available that facilitate Bayesian Network construction based on expert opinions. A subset of these tools also provides support for the ICI approach, including BayesiaLab, Norsys and Bayes Server. Among these this study uses BayesiaLab [24].

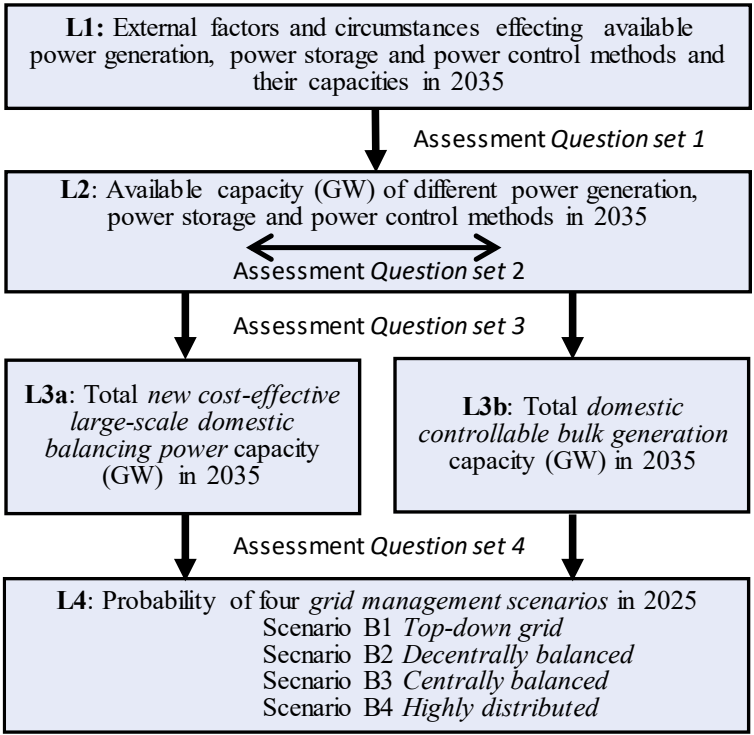

Fig. 3. Hierarchical structure of the Bayesian Network model and related four sets of assessment questions.

## 4. RESULTS

In Section 4.1, we present the literature review of the existing scenarios for the Finnish electricity system in 2035. Section 4.2 reports the views of the expert panel on the same topic. Finally, Section 4.3 introduces the developed hierarchical Bayesian Network model and presents the findings from the target optimisation analysis conducted with this model.

**4.1 Review of existing 2035 scenarios for the Finnish electricity system**

As mentioned already, multiple electrification scenarios have been devised in Finland for 2035, 2045 and 2050 to assesses alternatives for the power generation. These are summarised in Table 1, which includes a total of nine different scenarios: four by the Finnish Transmission System Operator (TSO) Fingrid [1], two by the Finnish Innovation Fund Sitra [2], and three by the Finnish Government [3]. The four scenarios by the Finnish Transmission System operator (TSO) Fingrid are called "*Power to products*", "*Hydrogen from wind*", "*Windy seas*", and "*Local power*". The scenarios from Finnish Innovation Fund Sitra are named "*Direct electrification scenario*" and "*Increased Power to X (P2X)*", while the three scenarios by the Finnish Government are "*Reference scenario*", "*Basic electrification scenario*", and "*Smart electrification scenario*". Table 1 also presents the current (early 2020s) power generation mix [25],[26],[27]. The scenarios from the three organisations use somewhat different categorisations for the power generation capacity mix. These are unified into three buckets: (1) production capacity, (2) control capacity consisting of import & export and new Demand Side Response (DSR) potential, and (3) new storage capacities. It is important to note that no distinct estimations are provided for nuclear and small-scale nuclear capacities, and only total capacity estimations (or none at all) are available, for storage methods. Referring to the terminology of Ref. [4] and Fig. 1, the *controllable domestic bulk generation consists* of large-scale and small-scale nuclear, biofuel, natural gas and (other) fossil fuel-based generation, as well as hydro power. These are highlighted by the green colour in Table 1. The *new cost-effective large-scale domestic balancing powers solutions* consist of new DSR potential, new storage capacity including pumped hydro storage, large-scale batteries, Power-to-X-to-Power (P2X-X2P) as well as home generation and batteries. These are highlighted by the blue colour in Table 1. Large-scale wind and large-scale solar power do not have any colour coding since they are variable, "non-controllable" as they are not necessarily able to produce any power even if the electricity demand would be high. Import does similarly not have any colour coding since it is not domestic and might not be an option even though there would high demand for electricity in Finland. The two columns on the right provide a summary of the *controllable domestic bulk generation* (green background colour) and the *new cost-effective large-scale domestic balancing power* (blue colour) in each scenario.

Table 1. Summary of the scenarios by some reputable organisations on the Finnish power generation mix capacities in 2035 and the early 2020s situation. Green background = *domestic controllable bulk generation*, blue background = new *cost-effective large-scale balancing power*, Governm = Finnish Government, DSR = Demand Side Response.

| Scenario | | Production capacity (in GW) | | | | | | | | | Control capacity (in GW) | | | New storage capacity (in GW) | | | | | Domestic controllable bulk generation (GW) | New cost-effective large-scale domestic balancing power (GW) |
|---|---|---|---|---|---|---|---|---|---|---|---|---|---|---|---|---|---|---|---|---|
| Scenario Name | Focus Year | Large-scale Nuclear | Small-scale nuclear | Large-scale wind | Large-scale solar | Bio | Gas | Fossil | Hydro | Sub-total | Import/ Export | New DSR potential | Sub-total | Pumped Hydro storage | Large-scale batteries | P2X-X2P | Home-gen & batteries | Sub-total | | |
| Fingrid 2023: Power to Products | 2035 | 4,0 | | 37,0 | 20,0 | | 4,0 | | 3,0 | 68,0 | 7,0 | 6,6 | 14,9 | 4,0 | | | | 4,0 | 11,0 | 10,6 |
| Fingrid 2023: Hydrogen from Wind | 2035 | 3,0 | | 44,0 | 15,0 | | 4,0 | | 2,0 | 68,0 | 6,0 | 6,2 | 13,5 | 4,0 | | | | 4,0 | 9,0 | 10,2 |
| Fingrid 2023: Windy Seas | 2035 | 4,0 | | 28,0 | 6,0 | | 3,0 | | 3,0 | 44,0 | 7,0 | 6,4 | 14,7 | 1,0 | | | | 1,0 | 10,0 | 7,4 |
| Fingrid 2023: Local Power | 2035 | 6,0 | | 14,0 | 7,0 | | 3,0 | | 3,0 | 33,0 | 6,0 | 4,2 | 11,5 | 3,0 | | | | 3,0 | 12,0 | 7,2 |
| Mean, Fingrid | 2035 | 4,3 | | 30,8 | 12,0 | | 3,5 | | 2,8 | 53,5 | 6,5 | 6,8 | 13,7 | 3,0 | | | | 3,0 | 10,5 | 8,9 |
| | | | | | | | | | | | | | | | | | | | | |
| Sitra 2021: Direct electrification | 2035 | 4,5 | | 24,0 | 1,5 | 3,0 | 4,5 | 2,5 | 3,2 | 43,2 | 5,2 | 17,1 | 23,6 | NA | 0,1 | 0,0 | NA | 0,1 | 17,7 | 17,2 |
| Sitra 2021: Increased P2X | 2035 | 4,5 | | 24,0 | 1,5 | 3,0 | 4,5 | 0,0 | 3,2 | 40,7 | 5,2 | 17,7 | 24,2 | NA | NA | 0,0 | NA | 0,0 | 15,2 | 17,7 |
| Mean, Sitra | 2035 | 4,5 | | 24,0 | 1,5 | 3,0 | 4,5 | 1,3 | 3,2 | 42,0 | 5,2 | 18,7 | 23,9 | NA | 0,1 | 0,0 | NA | 0,1 | 16,5 | 17,5 |
| | | | | | | | | | | | | | | | | | | | | |
| Governm 2021: Reference scenario | 2035 | 5,6 | | 7,0 | 3,0 | | 5,5 | | 3,0 | 24,1 | 5,7 | NA | NA | NA | NA | NA | NA | NA | 14,1 | NA |
| Governm 2021: Basic Electrification | 2035 | 5,6 | | 10,0 | 3,0 | | 6,1 | | 3,0 | 27,7 | 5,7 | NA | NA | NA | NA | NA | NA | NA | 14,7 | NA |
| Governm 2021: Smart Electrification | 2035 | 5,6 | | 11,0 | 4,0 | | 5,4 | | 3,0 | 29,0 | 5,7 | NA | NA | NA | NA | NA | NA | NA | 14,0 | NA |
| Mean, Finnish Government | 2035 | 5,6 | | 9,3 | 3,3 | | 5,7 | | 3 | 26,9 | 5,7 | NA | NA | NA | NA | NA | NA | NA | 14,3 | NA |
| | | | | | | | | | | | | | | | | | | | | |
| Min-Max ; all scenarios | 2035 | 3-6 | | 7-44 | 1,5-20 | | 3-10 | | 2-3,2 | 24,1-68 | 5,2-7 | 4,2-17,7 | 11,5-24,2 | 0-4 | | | | 0-4 | 9-17,7 | 7,2-17,7 |
| Mean Bulk and Balancing capacity | 2035 | | | | | | | | | | | | | | | | | | 13,8 | 13,2 |
| Current situation in Finland | 2023 | 4,4 | | 6,1 | 1,0 | | 6,0 | | 2,5 | 20,0 | 3,8 | 1,3 | 5,1 | 0 | | | | 0,0 | 12,9 | 1,3 |

Table 1 was presented to the experts as background information before the interviews. As noted in the bottom row, in early 2020s, the controllable bulk generation capacity is about 12,9 GW (including the Olkiluoto 3 nuclear power plant which started regular electricity production in 2023) and the balancing power capacity 1,3 GW [25],[26],[27]. By 2035, the mean capacities are projected to be 13,8 GW for *controllable domestic bulk generation* and 13,2 GW for *new large scale balancing power* (the second row from the bottom). However, these figures exhibit significant variability, as reflected in the wide range between the minimum and maximum values across different scenarios (cf. the row Min-Max; all scenarios). This is particularly attributable to the wind and solar energy capacities across the scenarios. Interestingly, noticeable increases in storage capacity can only be observed in Fingrid's scenarios probably because of high wind and solar capacities. As background information, the total electric energy consumption in Finland in 2023 was 79,8 TWh which is 14,2 MWh per capita [28].

## 4.2 Expert panel's assessment of the Finnish electricity system in 2035

This section reports the expert panel views on the Finnish electricity system in 2035 based on the conducted interviews applying the *Question Sets 1-4*, as presented in Chapter 3. Table 2 summarises the outcome of the *Question Set 1a*, which estimates power generation capacity mix in 2035. According to the survey results in Table 2, the experts predict that the overall *domestic controllable bulk generation* capacity will reach 12,7 GW in 2035, which would be comparable to current level of 12,9 GW (cf. Table 1). However, they anticipate significant increase in wind, solar, and DSR capacities, projecting them to roughly triple. The assessment of the power generation capacity mix is subject to considerable uncertainty, with confidence levels ranging between 49% and 72%, except for large-scale nuclear power and hydro power, which have more definitive assessments.

Table 2. The power generation capacity mix in 2035, experts not weighted vs. weighted. Green indicates bulk and blue balancing power.

| Power generation mix component | Non-weighted Capacity (GW) | Weighted Capacity (GW) | Confidence (%) |
|---|---|---|---|
| Large-Scale Nuclear | 4,7 | 5,0 | 85,2 |
| Hydro | 2,8 | 2,8 | 84,7 |
| Import/Export | 5,9 | 5,8 | 72,0 |
| Small-Scale Nuclear | 0,3 | 0,2 | 67,0 |
| Fossil | 0,6 | 0,5 | 66,7 |
| Wind | 19,1 | 19,0 | 66,3 |
| Solar | 5,8 | 5,9 | 65,8 |
| Battery | 1,2 | 1,1 | 61,3 |
| Pumped Hydro | 0,6 | 0,6 | 60,0 |
| DSR | 4,9 | 4,7 | 59,0 |
| Gas | 1,6 | 1,6 | 54,7 |
| Bio | 2,6 | 2,6 | 54,0 |
| Home | 0,9 | 0,8 | 54,0 |
| P2X-X2P | 0,7 | 0,6 | 49,3 |
| Total Capacity | 51,6 | 51,3 | |
| Total Controllable Bulk Capacity | 12,5 | 12,7 | |
| Total New Balancing Capacity | 7,5 | 7,1 | |

Fig. 4 presents the results from *Question Set 1c*, illustrating the causal effects of the highest-ranked external factors as well as the Leak, on various components of the power generation capacity mix. It is evident that policy and incentives, electricity pricing, and technology development, along with the Leak, play the most significant roles due to their high causal values and the number of components they influence. Conversely, factors such as geopolitical situation, solar irradiance, land resources, and system management volution have a minor impact on the power generation capacity mix. It is not surprising that, at a sum level the Leak significantly influences the components of the power generation capacity mix, given the multitude of other external factors that can impact the power generation capacity mix components. Policy and incentives have a substantial impact on the development of small-scale nuclear capacity, as it was listed 12 out of 15 panellists. This is expected, as both regulatory framework and public sentiment would need to evolve to permit the small-scale nuclear power plants closer to urban centres and residential areas. In fact, this shift is already ongoing in Finland, since the Radiation and Nuclear Safety Authority has in early 2024 abolished the requirement of fixed size precautionary action (protection) zones around nuclear power plants [29]. The significance of batteries is determined by the electricity pricing. The higher price volatility would likely be necessary to make the extensive use of relatively expensive battery technologies economically viable. Policy and incentives clearly also impact the use of fossil fuels. Without regulatory measures such as emission trading, these would economically be superior to many other generation types (when externality factors are not considered). The experts are sceptical about the P2X-X2P technologies: the maturation of the P2X-X2P technologies was mentioned by 11 panellists. This is possibly due to the lack of large-scale deployments and long-term operation of industrial size electrolysers and due to concerns on their efficiency.

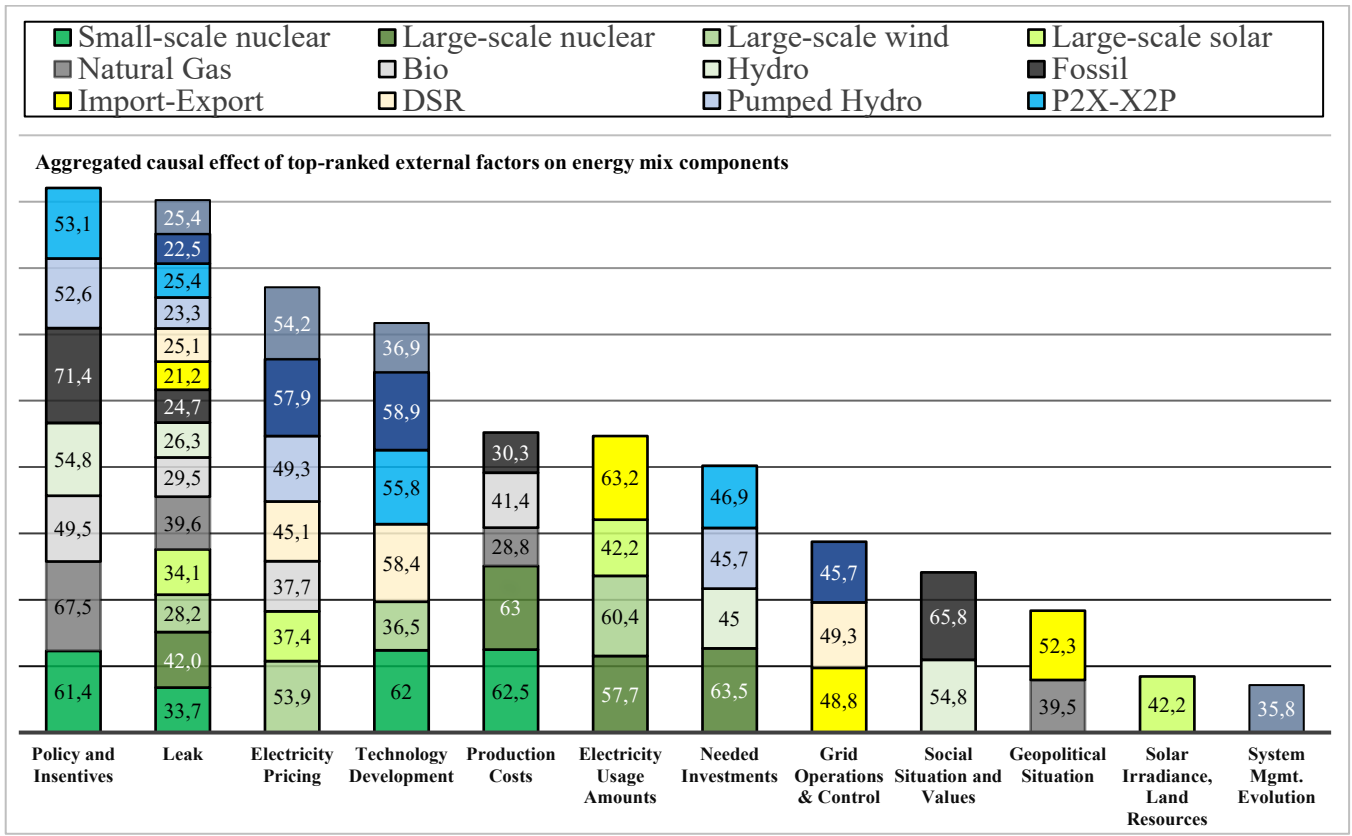

Fig. 4. Aggregated causal effect (y axis) of the top ranked external factors (x-axis) the Leak on different power generation capacity mix components (indicated by colour codes). The numbers, indicate the mean causal effect (as percentage figure) given by panellists for each energy mix component.

Fig. 5 provides an overview of the panel's views on the relationship between wind and solar capacities versus storage methods, covered in *Question Set 2*. The four rightmost bars show that direct and curtailed usage represents the highest consumption, ranging from 43% to 66% of total wind and solar production capacity. Notably, this portion of direct and curtailed usage decreases as wind and solar capacities increase. Conversely, in the P2X-X2P scenario, a similar but opposite trend is observed.

Fig. 6 summarises the causal effect of various power generation components on the total bulk capacity, assuming that the total *domestic controllable bulk generation capacity* in 2035 will be 13 GW or more (*Question Set 3a*). Hydro power and large-scale nuclear are identified as having the highest impact supported with the highest confidence values. In contrast, small-scale nuclear shows an even lower impact than the Leak, indicating the possibility of other potential energy generation sources or factors not covered in this study. Experts have proposed multiple reasons for the Leak: a reduced need for controllable bulk due to increased storage capacity, lower-than-expected electricity usage, and potential industry side streams for power generation.

Fig. 7 summarises the causal effect of various balancing power solutions on total balancing power capacity assuming that the 2035 total capacity of the *new cost-effective large-scale domestic balancing power capacity* will be 5 GW or more. Notably the Leak has the second highest impact after DSR. This suggests that, according to the experts, there may be other balancing power solutions or reasons not included in the analysis that could affect the total balancing power capacity. Based on the *Question Set 3b*, examples given by experts for missing solutions include gas turbines, integrated home solutions, and power import.

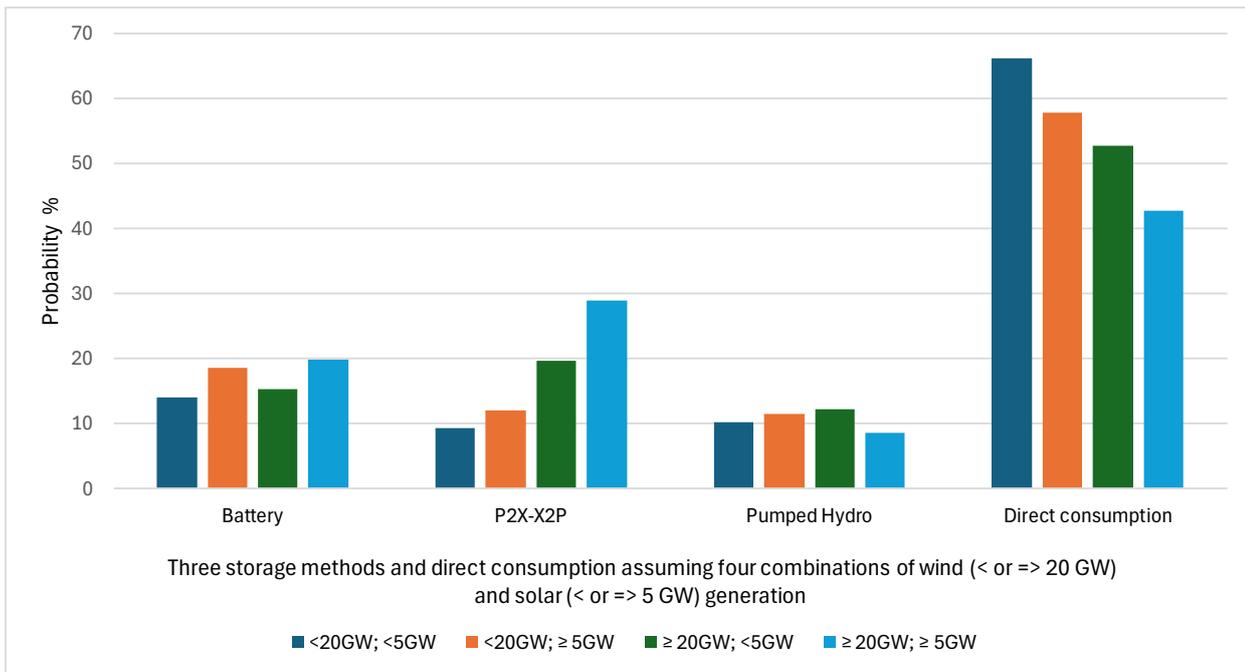

Figure 5. Relationship between wind and solar capacities vs. storage methods. The first number in each combination in the legend indicates the wind generation capacity, the latter solar capacity. The colour codes represent different capacity range combinations (wind below or above 20 GW, solar below or above 5 GW).

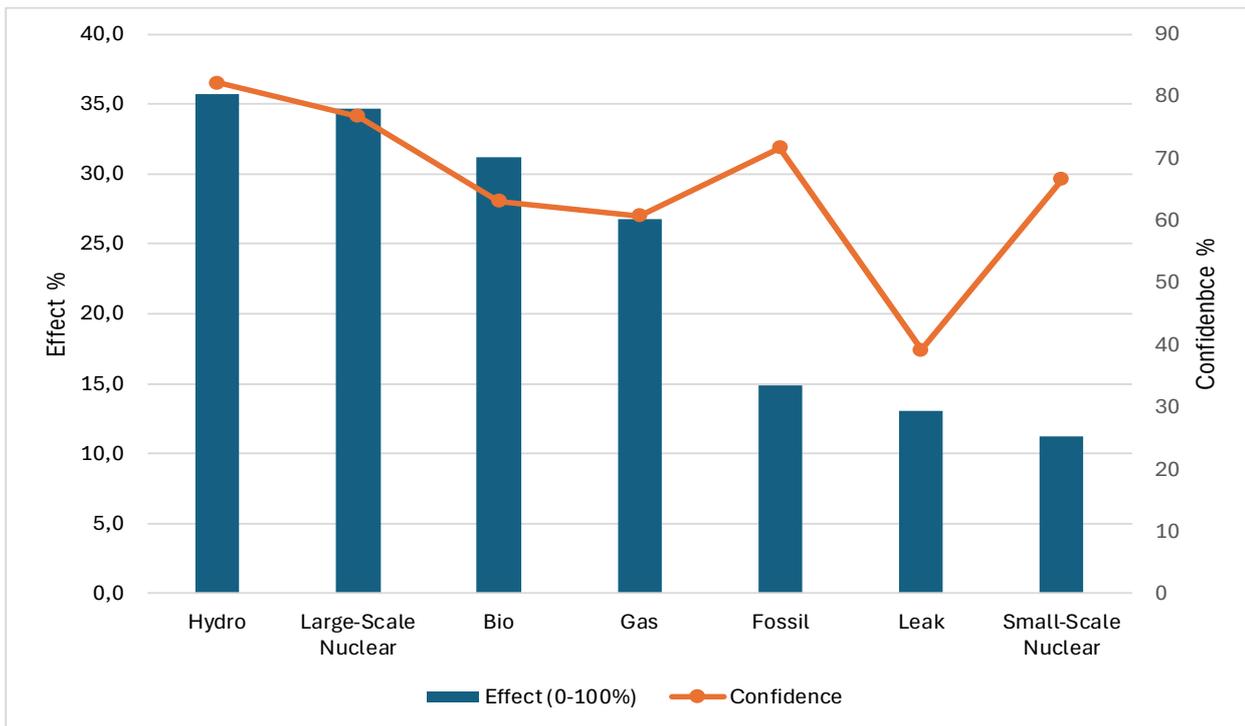

Fig. 6. Causal effect of various power generation capacity components on the total *domestic controllable bulk generation capacity* assuming the 2035 total domestic controllable bulk generation capacity will be equal or more than 13 GW.

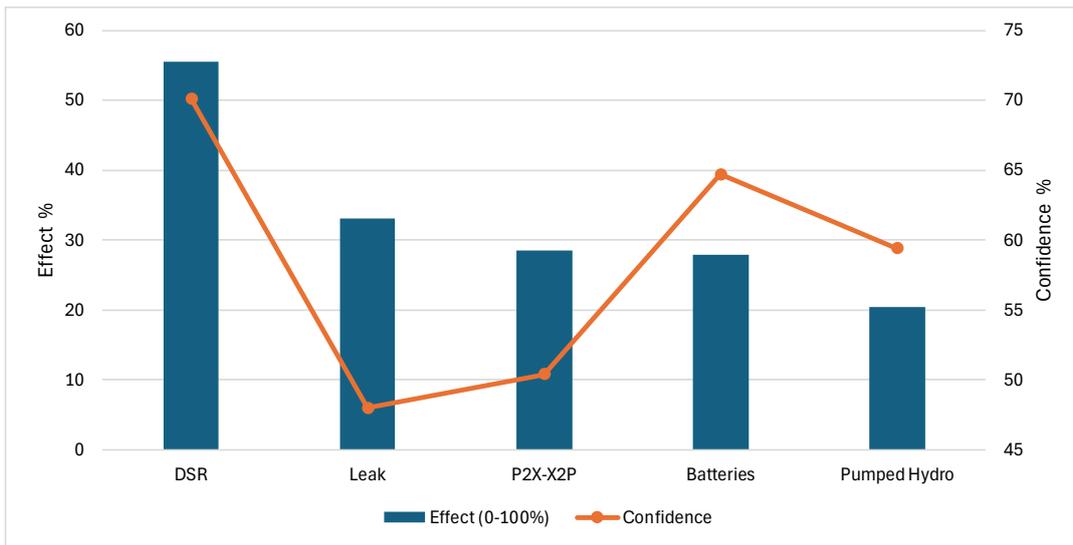

Fig. 7. Causal effect of various balancing power solutions on total balancing power capacity assuming that the 2035 total capacity of the *new cost-effective large-scale domestic balancing power* solutions will be 5 GW or more.

Chapter 1 and Ref [4] presented four *grid management scenarios*: Scenario B1 *Top-down grid*, Scenario B2 *Decentrally balanced grid*, Scenario B3 *Centrally balanced grid*, and Scenario B4 *Highly distributed grid*. Table 3 presents the likelihood of these scenarios based on the expert panel assessment with the *Question Set 4*. Not surprisingly, Scenario B4, *the Highly distributed grid*, has its highest likelihood (31,2%) when there is a lack of both *domestic controllable bulk generation* and *new cost-effective large-scale balancing power* solutions. Conversely, Scenario B1 *Top-down grid*, is the most likely scenario with a likelihood of 53,2% when these resources are abundant. Similarly, Scenario B3 *Centrally balanced grid*, has its highest likelihood (39,5%) when there is plenty of balancing power but less bulk power.

Table 3. Probability of *grid management scenarios* in 2035 as a function of different bulk and balancing power capacities.

| Bulk & Balance (GW) | Scenario B1 *Top-down grid* | Scenario B2 *Decentrally balanced grid* | Scenario B3 *Centrally balanced grid* | Scenario B4 *Highly distributed grid* |
|---|---|---|---|---|
| Bulk < 13; Balance < 5 | 22,9 | 16,5 | 29,4 | 31,2 |
| Bulk < 13; Balance ≥ 5 | 24,9 | 20,8 | 39,5 | 14,8 |
| Bulk ≥ 13; Balance < 5 | 31,0 | 26,2 | 30,3 | 12,5 |
| Bulk ≥ 13; Balance ≥ 5 | 53,2 | 11,9 | 26,7 | 8,2 |
| **Mean Probability** | **33,0** | **18,9** | **31,5** | **16,7** |

### 4.3. Bayesian Network construction and analysis

Fig. 8 illustrates the structure of the constructed hierarchical Bayesian Network. Nodes represent the model variables organised into four layers, as defined in Fig. 3 (layer 1 being on the top): (1) external factors, (2) power generation capacity mix components, (3) bulk and balancing power capacity, as well as (4) grid management scenarios. The arcs depict the causal relationships, showing how causes are linked to the effects. The nodes called ICI, OR, and the Leak form the Independence of Causal Influence (ICI) concept. ICI and the Leak nodes contain mean results from *Question Sets 1c* and *3*, while OR nodes multiply the individual results according to the Noisy-OR equation and form the 'parent friendly divorcing' structure. Furthermore, Fig. 8 shows the probability distributions of each node in this network excluding the probability distributions of the ICI nodes for clarity. Additionally, it includes an approximation of the actual capacity value (field "Value" inside the nodes).

As shown in Fig. 8, each network node is binary except for Grid Management, which has four values representing the probabilities for Scenarios B1-B4. The external factors are the highest-ranked nodes derived from *Question Set 1b* and described in Fig. 4 (excluding System Management Evolution due to its minor causal effect, cf. Fig. 4). Their True/False distribution appears even because the questions in *Question set 1a* and *1b* did not directly address it. However, True/False values have been assigned to best reflect their typical characteristics. For instance, the variable Needed Investments can be categorised as either "high" or "low". Each power generation mix node in the network (for example, Large-Scale Nuclear) has been categorised into two ranges: below the mean capacity in GW or greater than or equal to the mean capacity. The mean capacity is *a priori* power capacity mean from 15 experts, obtained from *Question Set 1a* (cf. Table 2). As described in Chapter 3, the total *domestic controllable bulk generation* capacity (Bulk Power in the Bayesian Network model) has been categorised as below 13 GW or greater than or equal to 13 GW. Similarly, the total capacity of *new cost-effective domestic large-large balancing power* solutions (Balancing Power in the Bayesian Network model) has been categorised as below 5 GW or greater than or equal to 5 GW.

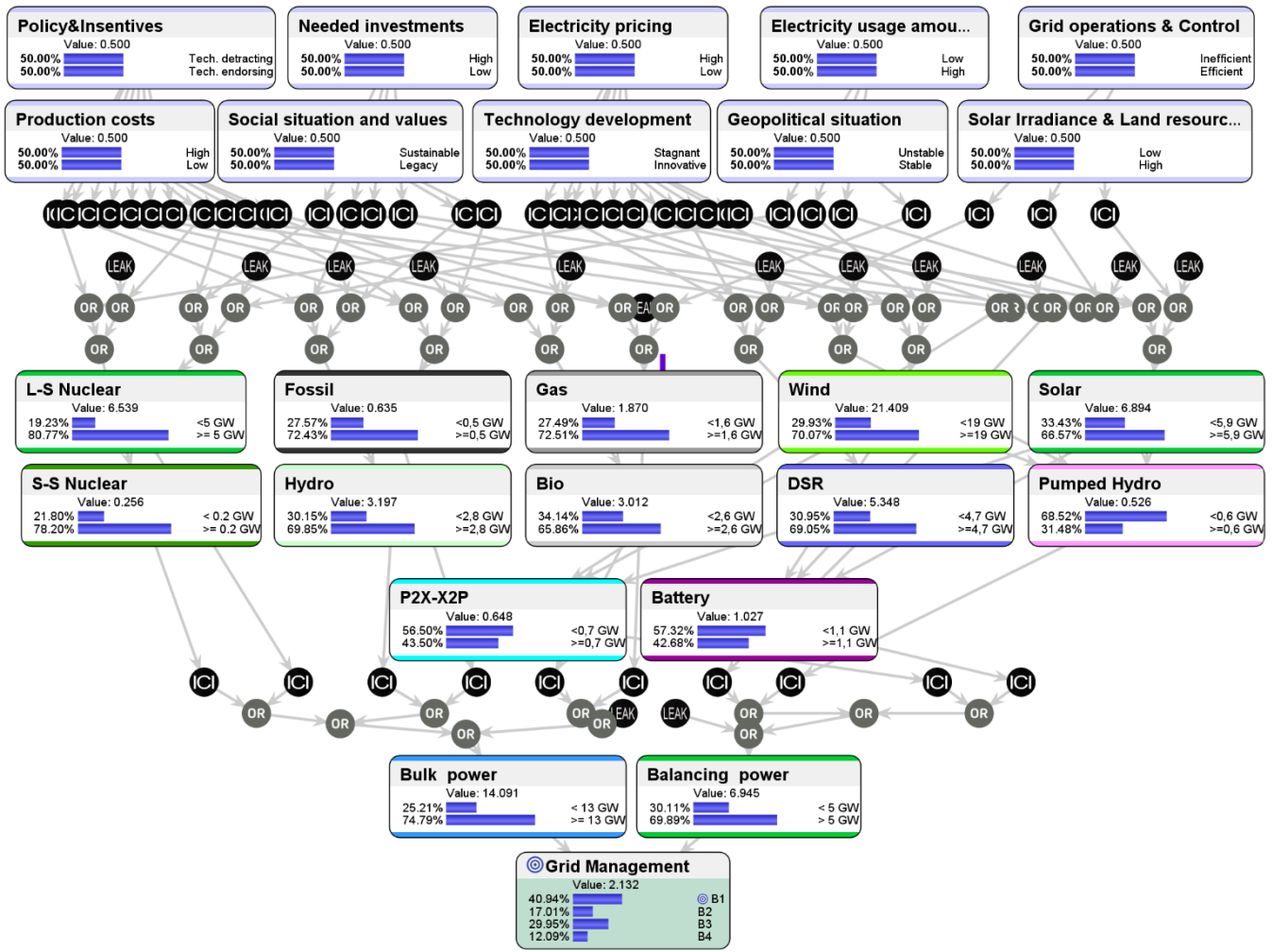

Fig. 8. The constructed hierarchical Bayesian Network. Bulk power = *controllable domestic bulk generation power*; Balancing power = *new cost-effective domestic large-large balancing power*; Grid management = *grid management scenarios*.

The probability distributions of layer L2-L4 nodes in Fig 8. represent *a posteriori* probabilities. These final probability distributions for power generation capacity mix components, as well as for total *domestic* *controllable bulk generation* and *new cost-effective domestic large-large balancing power* capacities, are outcomes of the ICI process described in Chapters 2 and 3 and are indicated in the nodes of Fig. 8. In Bayesian statistics *a posteriori* distributions differ from the *a priori* distributions due to the integration of new evidence from expert assessments with ICI approach. For instance, the probability that large-scale nuclear capacity in 2035 will be greater than or equal to 5 GW is 80,8% (cf. node L-S Nuclear in Fig. 8) instead of 50%, and the probability that P2X-X2P capacity in 2035 will be less than 0,7 GW is 56,5% (c.f. node P2X-X2P in Fig. 8) instead of 50%. Similarly, the probability that the total *domestic controllable bulk generation* capacity in 2035 will be greater than or equal to 13 GW is 74,8%, and the probability that the total *new cost-effective domestic large-large balancing power* capacity will be greater than or equal to 5 GW is 69,9% instead of 50%. Similarly, compared to the *a priori* distributions regarding the probabilities for the *grid management scenarios* (cf. the mean values in Table 3), *a posteriori* distributions for Scenarios B1-B4 are 40,9%, 17%, 30%, and 12,1%, respectively, instead of 33%, 18,9%, 31,5% and 16,7% (cf. Table 3).

As mentioned previously, each node in Fig. 8, contains not only a probability distribution but also an approximation of the actual capacity value. This approximation is based on the assumed Gaussian distribution of numerical estimates provided by the experts around the calculated mean capacity. Expert estimates have been categorised into two groups: those above the overall mean and those below it, with sub-means calculated for each bucket. The approximated values are then obtained by multiplying the probability distribution values by the sub-means. For example, in the case of Large-Scale Nuclear power, experts' estimations are first divided into two groups: those below the mean of 5 GW and those greater than or equal to 5 GW. The means of these groups are then calculated, resulting in 2,5 GW for the below 5 GW and 7,5 GW for the greater or equal to 5 GW. In this way, if the probability distribution is evenly split at 50%, the resulting value for Large-Scale Nuclear power is 5 GW. If the probability distribution is entirely (100%) for capacities greater than or equal to 5 GW, then the resulting capacity value is 7,5 GW.

Table 4 demonstrates that integrating evidence of external factors through the ICI process adjusts the capacities of various power generation capacity mix components, creating *a posteriori* capacities. The table shows the initial mean expert estimations (*a priori* assessment) for power generation capacity mix components in 2035 and the revised estimations following evidence of the effect of external factors on the power generation capacity mix components through the ICI approach and constructed Bayesian Network (*a posteriori* estimations in Fig. 8). As seen from the Table 4, Wind, Large-Scale Nuclear, Solar, and DSR capacities increase by 2,4 GW, 1,5 GW, 1 GW, and 0,7 GW, respectively, while the capacities of pumped hydro and batteries slightly decrease.

Table 4. *A priori* capacity estimations of power generation capacity mix components based on the *Question Set 1a,* and *a posteriori* estimations after having incorporated *Question Set 1c* results to the constructed Bayesian Network with ICI -approach.

| Energy Mix Component $X_1$ | *A priori* Capacity Estimations (GW) | *A posteriori* Capacity Estimations (GW) | Delta (GW) |
|---|---|---|---|
| Large-Scale Nuclear | 5,0 | 6,5 | 1,5 |
| Hydro | 2,8 | 3,2 | 0,4 |
| Small-Scale Nuclear | 0,2 | 0,3 | 0,1 |
| Fossil | 0,5 | 0,6 | 0,1 |
| Wind | 19,0 | 21,4 | 2,4 |
| Solar | 5,9 | 6,9 | 1,0 |
| Battery | 1,1 | 1,0 | -0,1 |
| Pumped Hydro | 0,6 | 0,5 | -0,1 |
| DSR | 4,7 | 5,4 | 0,7 |
| Gas | 1,6 | 1,9 | 0,3 |
| Bio | 2,6 | 3,0 | 0,4 |
| P2X-X2P | 0,6 | 0,7 | 0,1 |
| **Total Capacity** | **44,7** | **51,4** | **6,7** |

Similarly, the *a posteriori* capacities for total *domestic controllable bulk generation* and total *new cost-effective domestic large-scale balancing power* solutions can be estimated based on the new evidence using the constructed Bayesian Network and the methods described in Chapters 2 and 3. Since experts were not asked to estimate the total capacities directly, but rather to estimate how individual power generation capacity mix components affect the total *domestic controllable bulk generation* and *new cost-effective domestic large-scale balancing power* capacities as per the *Question Sets 3a* and *3b*, the results from Table 1 were first used to approximate the proper *a priori* capacities for *domestic controllable bulk generation* and *new cost-effective domestic large-scale balancing power* capacities in 2035 (i.e., 13 GW for bulk and 5 GW for balancing capacity, the so called trigger values explained in Chapter 3). With help of the constructed Bayesian Network, the *a posteriori* capacity for the *total domestic controllable bulk generation* is estimated at 14,1 GW (cf. Bulk power in Fig. 8) and for the total *new cost-effective domestic large-scale balancing generation* 7,0 GW (cf. node Balancing power in Fig. 8).

The constructed Bayesian Network was utilised to conduct target optimisation aiming at maximising the probability of the Scenario B1 *Top-down grid*. Scenario B1 was selected for the Target Optimisation since, as described in Chapter 1, in Scenario B1 both *controllable bulk domestic generation* and *new cost-effective domestic large-scale balancing power* capacity would be roughly adequate. Thus, Scenario B1 does not necessitate large-scale DSR and storage solutions and extensive automatisation for maintaining grid stability. Additionally, in situations where import capacity is limited, Scenario B1 is the preferred one. In the target optimisation analysis, greedy search is used to identify the optimal capacities (GW) for each power generation capacity mix component in descending order to maximise or minimise the probability of the Scenario B1, the target.

The probability of Scenario B1 *Top-down grid* is maximised by applying the following equation for each power generation capacity mix component:

**Score** = $w_1$ x $I_i$ x $w_2$ x $E_i$ x $w_3$ x 1 /$C_i$ \hfill Eq. 2

where:
- **Score** is the value which will be maximised
- $I_i$ is the impact (effect) size (%) of an power generation capacity mix component $X_i$ in the Bayesian Network, measured as a delta between *a posteriori* and *a priori* probabilities of the Scenario B1, when $X_i$ has the proposed *a posteriori* value $x$ found based on the greedy search compared to its *a priori* state.
- $E_i$ is the evidence, calculated as the joint probability (%) of the Bayesian Network when $X_i = x$
- $C_i$ represents the construction cost to build one GW of energy for power generation capacity mix component $X_i$.
- $w_1$, $w_2$ and $w_3$ are weights, each set to a value of 1 in this study.

According to Eq. 2, both a higher impact size and joint probability result in a higher score. Similarly, the lower the cost, the higher the score. The Eq. 2 is applied iteratively by optimising one energy power generation capacity mix component $X_i$ at a time to maximise the score. After finding the component and its value which maximises the score, the evidence for $X_i$ is updated in the Bayesian Network model and the score is re-maximised using the rest of the power generation capacity mix components until all of them are covered. Thus, Eq. 2 provides a prioritised list of power generation capacity mix components, ranked from most to least important, along with their proposed capacity values that lead towards the optimal goal.

Table 5 summarises the results from the target optimisation. It shows the prioritised list of power generation capacity mix components from the most important (DSR) to the least important, along with their proposed new capacity values. As seen, applying Eq. 2 would increase the probability of the Scenario B1 from the original 40,9 % to 46 ,9% and DSR, gas power and batteries are the three most important and large-scale nuclear power, P2X-X2P and small-scale nuclear power the least important power generation capacity mix components in reaching the maximum Scenario B1 probability. The construction costs (overnight costs) costs are derived from Ref. [30] with the exception pumped hydro, which is based on Ref [31]. Although these costs are largely based on the U.S. averages, they offer indicative and directionally relevant benchmarks also for our purposes. Overnight costs are used in the energy industry to provide quick, simplistic comparisons of building power plants [32]. They exclude factors like interest rates and life spans of plants. Since reliable consensus sources for DSR and P2X-X2P costs were unavailable, these estimates were provided by the authors. DSR deviates from the other technologies listed in Table 5 since it is not a power generation technology. If there is suitable consumption that can and is willing to be adjusted down according to power grid needs, the implementation of DSR capacity could be fairly inexpensive; otherwise, its cost could be prohibitively high. Finland having currently peak power of about 15 GW [33] and committed DSR capacity of 1,3 GW (c.f. Table 1), additional DSR capacity of several GWs is not possible – people and industries cannot or are not willing to curtail their consumption essentially more during the coldest winter days. Therefore, an increase of DSR capacity would require new DSR applications, such as electrified district heating with large thermal storage or hydrogen production plants. The cost estimate for DSR capacity in Table 5 is set at the lowest value, i.e., 800 million dollars per GW. P2X-X2P differs from the other generation technologies since it has a significant technology risk due to so far lacking base of large commercial deployments. For example, as of today, the largest electrolysers being deployed in Europe consume some tens of MWs of electrical power [34]. Due to technology risk and essential conversion losses, the cost estimate for P2X-X2P capacity is set at the highest value in Table 5, at 9.000 million dollars per GW.

Table 5. Target optimisation: A prioritised list of power generation capacity mix components, including their initial (see Table 2) and new proposed capacities, to maximise the probability of Scenario B1 *Top-down grid*.

| Power generation capacity mix component | *A priori* Capacity (GW) | *A posteriori* (Proposed) Capacity (GW) | Delta *a posteriori – a priori* (GW) | Costs per GW (Million $) | Joint Probability | Effect Size on B1 | Probability of B1 |
|---|---|---|---|---|---|---|---|
| *Starting point* | | | | | 50,0 % | | 40,9 % |
| DSR | 5,3 | 6,4 | 1,1 | 800 | 34,5 % | 1,6 % | 42,5 % |
| Gas | 1,9 | 2,2 | 0,3 | 867 | 25,5 % | 0,8 % | 43,3 % |
| Battery | 1,0 | 1,6 | 0,6 | 1270 | 11,2 % | 0,7 % | 44,0 % |
| Hydro | 3,2 | 3,8 | 0,6 | 3421 | 8,1 % | 0,9 % | 44,9 % |
| Bio | 3,0 | 3,9 | 0,9 | 4998 | 5,8 % | 0,5 % | 45,4 % |
| Pumped Hydro | 0,5 | 0,8 | 0,3 | 2202 | 2,0 % | 0,5 % | 45,9 % |
| Fossil | 0,6 | 0,8 | 0,2 | 2240 | 1,7 % | 0,1 % | 46,0 % |
| Wind | 21,4 | 25,0 | 3,6 | 2098 | 1,3 % | 0,1 % | 46,1 % |
| Solar | 6,9 | 8,9 | 2,0 | 1448 | 0,9 % | 0,1 % | 46,2 % |
| L-S Nuclear | 6,5 | 7,5 | 1,0 | 7777 | 0,7 % | 0,3 % | 46,5 % |
| P2X-X2P | 0,6 | 1,1 | 0,5 | 9000 | 0,5 % | 0,3 % | 46,8 % |
| S-S Nuclear | 0,3 | 0,3 | 0,0 | 8349 | 0,4 % | 0,1 % | 46,9 % |

## 5. DISCUSSION AND CONCLUSIONS

Table 6 summarises projected availability of generation capacities during the peak hour and peak season based on the *a posteriori* capacity estimations in Table 4. The peak season refers to an extended period of high electricity consumption, typically occurring during Finland's cold, dark, and low-wind winter weeks. For peak hour availability, assumptions include 6% availability of installed wind capacity (as per Ref. [26]), 0% for solar due to the low irradiance in Finland during winter, and 90% for other generation types. For example, in the United States the availability of nuclear power plants was 93% in 2023 [35]. For peak season, the availability assumptions are largely similar to those for peak hour but exclude battery, pumped hydro, and P2X-X2P storage solutions, which might lack the capacity for sustained output over several weeks. Battery limitations relate to energy density and physical size constraints, pumped hydro to the scarcity of large reservoirs and significant elevation changes, and P2X-X2P to the anticipated lack of large-scale deployments (cf. Section 4.3). As indicated in Table 6, the available generation during peak hour and during peak season would amount to 17,2 GW and during peak hour to 15,2 GW. By adding the anticipated import capacity of 5,8 GW (Table 2) to the peak hour and peak season capacities shown in Table 6, and assuming that energy imports are available from Sweden, Norway, and Estonia, the total capacity would reach 23 GW (17,2 + 5,8) for the peak hour and 21 GW (15,2 + 5,8) for the peak season. The peak power consumption in the Finnish power grid winter 2023-2024 occurred on the 3rd of January 2024 and was 15,0 GW [33]. While comparing this to the anticipated peak hour capacity of 23 GW, the increase from would be 8GW i.e. 53%. This increase could enable major additional electrification of the society by 2035. During peak hours, DSR solutions can help to even out the load. As can be seen in Table 2, the DSR capacity is expected to increase significantly from the current 1,3 GW by 4,7 GW. One reason to this result could be the electrification of the central heating system and the introduction of P2X technologies. These could potentially be switched off for some time, e.g. during the peak hour, without major impact. The DSR solutions are not equally effective during the peak season, as switching the heating of central heating water tanks or major industrial activities for weeks would not be feasible.

The expert projections clearly indicate a substantial increase of wind generation capacity by 2035, with capacity expected to rise by 14,5 GW, from 6,9 GW at the end of 2023 [36] to 21,4 GW (as shown in Tables 4 and 6), despite the limited capacity factor associated with wind-energy and even though the increase of seasonal electric energy storage such as pumped hydro and P2X-X2P would be fairly modest (Fig. 5). This substantial growth may be partly due to sector integration efforts, such as the anticipated electrification of central heating systems. As mentioned previously, the possibilities to build pumped hydro in Finland are limited and low figures for P2X-X2P might be attributable to the inefficiencies of the process and the current immaturity of the technology (cf. discussion in Section 4.3.) [37].

Table 6. Availability of the *a posteriori* generation capacity during peak hour and peak season

| Type of generation | *A posteriori* capacity 2035 from Table 4 (GW) | Availability during peak hour (GW) | Availability during peak season (GW) |
|---|---|---|---|
| **Large-scale Nuclear** | 6,5 | 5,9 | 5,9 |
| **Hydro** | 3,2 | 2,9 | 2,9 |
| **Small-scale Nuclear** | 0,3 | 0,3 | 0,3 |
| **Fossil** | 0,6 | 0,5 | 0,5 |
| **Wind** | 21,4 | 1,3 | 1,3 |
| **Solar** | 6,9 | 0,0 | 0,0 |
| **Battery** | 1,0 | 0,9 | 0,0 |
| **Pumped Hydro** | 0,5 | 0,5 | 0,0 |
| **Gas** | 1,9 | 1,7 | 1,7 |
| **Bio** | 3,0 | 2,7 | 2,7 |
| **P2X-X2P** | 0,7 | 0,6 | 0,0 |
| **Total Capacity** | 46,7 | 17,2 | 15,2 |

The results of the target optimisation (Table 5) reveal that the policy makers and industry stakeholders should prioritise increasing demand response, gas power, and battery storage capacities. One possible reason could be that these mature and controllable technologies facilitate large-scale integration of renewable wind and solar generation as they can be utilised to guarantee energy adequacy during peak consumption hours and seasons. Specifically, the target optimisation model recommends expanding capacities for DSR by 1,1 GW, for natural gas by 0,3 GW, for batteries by 0,6 GW, for hydro power by 0,6 GW, and bioenergy by 0,9 GW. Another interpretation of the target optimisation results is that in order to maximise the likelihood of the Scenario B1 *Top-down grid*, about 3.000 MUSD (800+867+1.270) should be invested in DSR, natural gas and batteries. Even though natural gas is a fossil fuel as well, its usage could be motivated for short periods to achieve overall economic feasibility of the energy system.

Bayesian Networks offer excellent possibilities for extensive sensitivity analysis, enabling researchers to model scenarios and assess the impact of uncertainties. The novelty of this study lies in utilising a sophisticated four-layer Bayesian Network and the related Independence of Causal Influence (ICI) approach to systematically capture and model expert opinions across four distinct question sets, encompassing a total of 64 questions, each accompanied by confidence estimates. The hierarchical structure was predefined by the authors to reflect hypothesised potentially interesting relationships. It facilitated the discussion and analysis by visualising the causal relationships. The use of ICI method was central in reducing the complexity of the analysis and the number of expert questions to a manageable level, as well as for simplifying the documentation of otherwise complex causal relationships. To the best of our knowledge, this study represents the most extensive application of the ICI method across any field. Although the scenarios with their probabilities and likelihoods are naturally influenced by the specificities of Finland's national context, the developed Bayesian Network framework and associated questions sets constitute broadly and readily applicable generic tools that can be directly utilised in exploring the causality of power grid evolution in other countries.

## LIST OF ABBREVIATIONS

| BN | Bayesian Network |
| --- | --- |
| CPT | Conditional Probability Table |
| DSR | Demand Side Response |
| ICI | Independence of Causal Influence |
| NGT | Nominal Group Technique |
| P2X | Power-to-X |
| RAM | RAND/UCLA Appropriateness Method |
| RAND/UCLA | RAND research organization / University of California at Los Angeles |
| TSO | Transmission System Operator |
| X2P | X-to-Power |

## ACKNOWLEDEGEMENTS

This work was supported in part by the Finnish public funding agency for research, Business Finland under the projects IFORGE (grant number 7127/31/2021) and SmartGrid 2.0 (grant number 4963/31/2022).

# APPENDIX A   The Question Sets

The questions regarding peak power capacities at hierarchy layer L2, as well as the impact of various variables at hierarchy layer L1 on these capacities, comprise Question Set 1. It consists of three distinct sub-questions: *Question Set 1a,1b and 1c*.

- **Question Set 1a**: What is the estimated projected capacity $Y_i$ in GW of $X_i$, where $X_i$ is one of the 14 capacity items in L2 in 2035? How confident are you in this estimation?
- **Question Set 1b**: Considering the 15 external factors in L1, please identify the three factors in random order that come to mind, which have the greatest impact on the capacity each individual item $X_i$ in layer L2.
- **Question Set 1c**: What is the causal effect of the three external factors obtained from Question Set 1b and the Leak on each of the 14 power generation capacity mix components $X_i$, if they were the only causes? How confident are you in the estimation.

A standard CPT approach is employed in Question Set 2 to estimate the dependencies between two parent nodes at L2 (wind and solar capacities in 2035) and a child node at L2.

- **Question Set 2**: What are the probabilities for the following scenarios, where their sum is 100%: (1) electricity used for large-scale batteries, (2) electricity used for P2X-X2P, (3) electricity used for pumped hydro storage, and (4) electricity used for direct consumption? These probabilities should be related to the four Parent1 - Parent2 value pairs 1. Wind: <20 GW - Solar:<5 GW, 2. Wind: <20 GW - Solar: ≥ 5GW, 3. Wind: ≥20 GW- Solar:<5 GW, 4. Wind: ≥20 GW- Solar: ≥ 5GW. How confident are you in this estimation?

To examine the causal effect of power generation capacity mix components in layer L2 on the two variables in L3 *with Question Set 3*, the ICI approach is used.

- **Question Set 3a**: Assuming the 2035 total capacity of the new cost-effective domestic large-scale balancing power solutions will be ≥5 GW, what would be the independent causal effect (as probability value between 0 – 100%) of each balancing solution $X_{bai}$ including the potential Leak on the given total balancing capacity if it were the only cause? If the Leak has been > 0%, what are the potential reasons for it.  How confident are you in these estimations?
- **Question Set 3b**: Assuming the 2035 total domestic controllable bulk generation capacity will be ≥13 GW, what would be the independent causal effect (as probability value between 0 – 100%) of each bulk energy source $X_{bui}$ including the potential Leak on the given total bulk capacity if it were the only cause? If the Leak has been > 0%, what are the potential reasons for it? How confident are you in these estimations?

Finally, the **Question Set 4** addresses the dependency between L3 and L4 in 2035, estimated as a probabilistic relationship between two parent variables and the four potential *grid management scenarios*, which acts as a child node.

- **Question Set 4**: What are the probabilities for the following scenarios, where their sum is 100%: Scenario B1Top-Down grid, Scenario B2 Decentrally balanced grid, Scenario B3 Centrally balanced grid  and Scenario B4 Highly distributed grid? These probabilities should be related to the four Parent1 - Parent2 value pairs: 1. Bulk <13 GW - Balance < 5 GW, 2. Bulk < 13 GW - Balance ≥ 5 GW, 3. Bulk ≥ 13 GW- Balance < 5 GW, 4 Bulk ≥ 13 GW - Balance ≥ 5 GW. How confident are you in this estimation?